\documentclass[aps,pra,reprint,superscriptaddress,showpacs,letterpaper,longbibliography]{revtex4-1}
\usepackage[english]{babel}
\usepackage{ucs}
\usepackage[utf8x]{inputenc}
\usepackage{amsmath}
\usepackage{amsfonts}
\usepackage{amssymb}
\usepackage{graphicx}
\usepackage{bm}
\usepackage{bbm}
\usepackage{empheq}
\usepackage{color}

\makeatletter
\@ifundefined{textcolor}{}
{%
 \definecolor{BLACK}{gray}{0}
 \definecolor{WHITE}{gray}{1}
 \definecolor{RED}{rgb}{1,0,0}
 \definecolor{GREEN}{rgb}{0,1,0}
 \definecolor{BLUE}{rgb}{0,0,1}
 \definecolor{CYAN}{cmyk}{1,0,0,0}
 \definecolor{MAGENTA}{cmyk}{0,1,0,0}
 \definecolor{YELLOW}{cmyk}{0,0,1,0}
}

\definecolor{yyl}{rgb}{.8,.05,.08}

\definecolor{thk}{rgb}{.05,.05,.8}

\begin{document}

\title{Phase locking of a semiconductor double quantum dot single atom maser}

\author{Y.-Y. Liu}
\author{T. Hartke}
\author{J. Stehlik}
\author{J. R. Petta}
\affiliation{Department of Physics, Princeton University, Princeton, New Jersey 08544, USA}

\date{\today}

\begin{abstract}
 
We experimentally study the phase stabilization of a semiconductor double quantum dot (DQD) single atom maser by injection locking.  A voltage-biased DQD serves as an electrically tunable microwave frequency gain medium. The statistics of the maser output field demonstrate that the maser can be phase locked to an external cavity drive, with a resulting phase noise $\mathcal{L}$ = -99~dBc/Hz at a frequency offset of 1.3 MHz. The injection locking range, and the phase of the maser output relative to the injection locking input tone are in good agreement with Adler's theory. Furthermore, the electrically tunable DQD energy level structure allows us to rapidly switch the gain medium on and off, resulting in an emission spectrum that resembles a frequency comb. The free running frequency comb linewidth is $\approx$ 8~kHz and can be improved to less than 1 Hz by operating the comb in the injection locked regime.

\end{abstract}

\pacs{73.21.La, 73.23.Hk, 84.40.lk}

\maketitle

\section{Introduction}

Narrow linewidth lasers have a wide range of applications in communication technology, industrial manufacturing, and metrology. \cite{Kleppner1962, Kleppner65, Diddams2000}. Unlike in atomic systems, where linewidths can approach 1 mHz \cite{Meiser2009, Kessler2012, Bohnet2012},  charge noise in semiconductor lasers typically leads to linewidths that are 10--100 times larger than the Schawlow and Townes (ST) prediction \cite{Schawlow1958, Siegman64, Milonni1988, Dick91, Bourgeois05, Oxborrow12}. It is therefore often desirable to stabilize the frequency of solid state masers/lasers using existing narrow linewidth sources via the injection locking effect \cite{Stover1966, Schunemann1998}. To achieve an injection-locked state, an external cavity drive is applied to the laser, resulting in stimulated emission at the frequency of the injected signal and a corresponding reduction in linewidth \cite{Adler1946}. In addition to frequency stabilization, the precisely locked phase can be used as a resource for other metrology applications. For example, the phase of an injection-locked, trapped-ion-phonon laser has been proposed for applications in mass spectrometry and as an atomic-scale force probe \cite{Herrmann2011}.  

In this paper we examine phase locking of a DQD semiconductor single atom maser (SeSAM) \cite{Liu2017}. Driven by single electron tunneling events between discrete zero-dimensional electronic states, this device results in microwave frequency photon emission with a free-running emission linewidth of 6 kHz. Due to low frequency charge noise, the linewidth is still 50 times larger than the ST limit \cite{Schawlow1958, Milonni1988, Liu2015, Liu2017}. Here we use injection locking to significantly improve the performance of the SeSAM. In contrast with our previous work, which demonstrated injection locking of a multi-emitter maser, we directly measure the degree of phase stabilization in the injection locked state by examining the photon statistics of the output field \cite{LiuPRA2015}. The locked maser output  achieves a phase noise better than $\mathcal{L}$ = -99~dBc/Hz (1.3 MHz offset). The locking phase and locking range are shown to be in good agreement with Adler's prediction \cite{Adler1946}. 

Looking beyond single-tone narrow linewidth sources, the electrically tunable energy level structure of the SeSAM allows the gain medium to be switched on and off. We explore the output of the SeSAM in both free running and injection locked modes while the DQD energy levels are periodically modulated at frequency $f_{\epsilon}$ \cite{Cassidy2017}. When the SeSAM is unlocked, it outputs a frequency comb with a mode spacing of $f_{\epsilon}$ and a 8 kHz linewidth. Under injection locking conditions, the linewidth of the modulated SeSAM frequency comb emission peaks is reduced to less than 1 Hz. These measurements demonstrate that a single cavity-coupled DQD may serve as a compact, low temperature microwave source that is suitable for use in quantum computing experiments. 

\section{Double Quantum Dot Micromaser}

\vspace{0.2cm}
\begin{figure}[t]
	\begin{center}
		\includegraphics[width=\columnwidth]{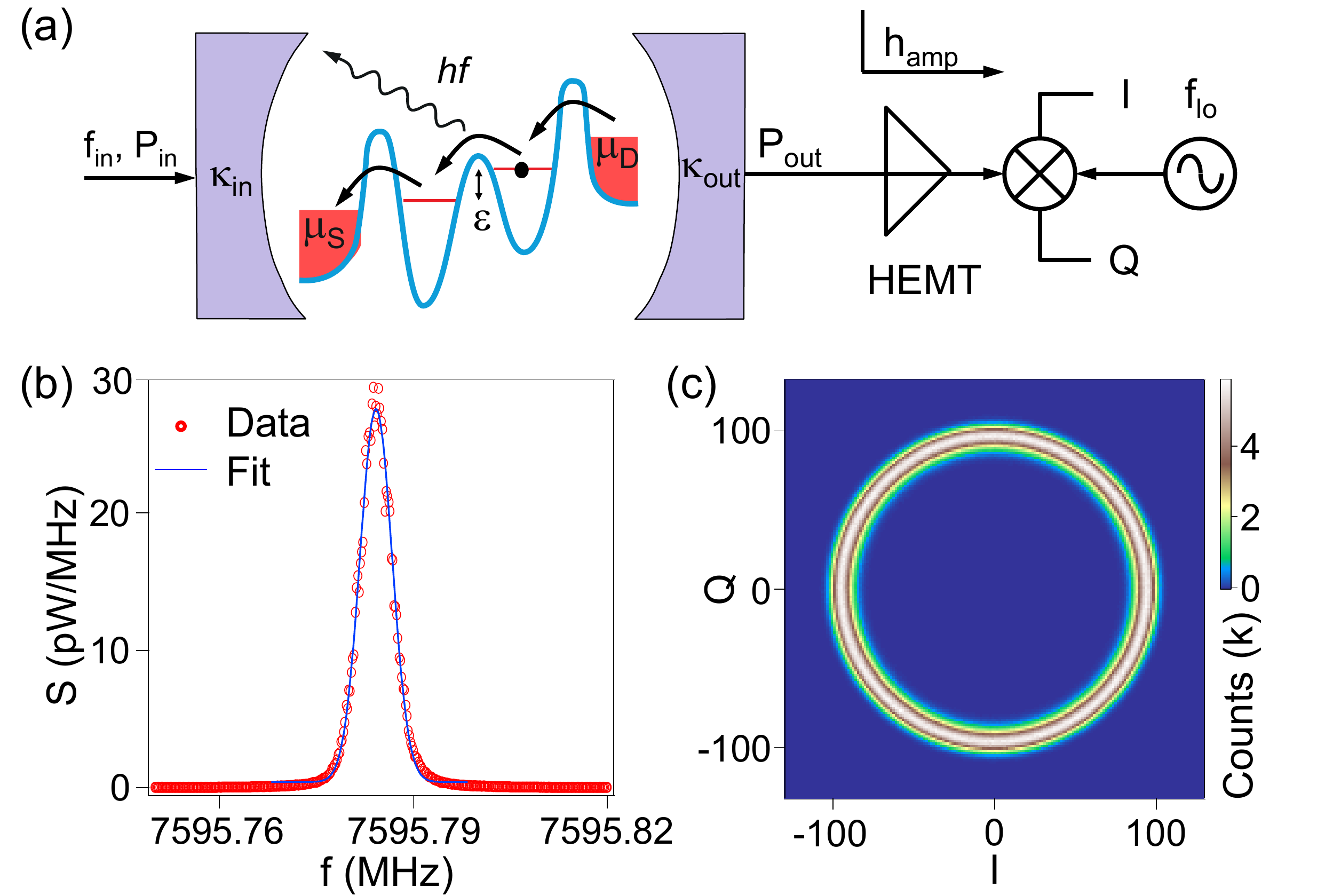}
		\caption{\label{Fig: setup} (a) Schematic of the experimental setup. The SeSAM consists of a DQD gain medium that is placed in a superconducting cavity. A source-drain bias voltage $V_{\rm SD}$ is applied to drive a current through the DQD and generate the microwave frequency photons that result in masing. (b) Free running ($P_{\rm in}=0$) power spectral density $S(f)$ of the maser output field. The data (circles) are fit with a Gaussian (solid line) with a FWHM $\Gamma$ = 5.6 kHz. (c) $IQ$ histogram of the maser output field measured with $P_{\rm in}=0$. The donut shape is indicative of above-threshold maser action.}
	\end{center}	
\end{figure}

The SeSAM is implemented in the circuit quantum electrodynamics architecture (cQED), where strong coupling has been demonstrated between microwave photons and a variety of mesoscopic devices \cite{Wallraff2004, Mi2017, Stockklauser2017, Bruhat2016}. As illustrated in Fig.~\ref{Fig: setup}(a), the maser consists of a single semiconductor DQD that is coupled to a microwave cavity \cite{Liu2017}. The DQD gain medium is formed from a single InAs nanowire that is bottom gated to create an electrically tunable double-well confinement potential \cite{Nadj-Perge2010, Wiel2002}. The DQD energy level detuning $\epsilon$ is gate-voltage-controlled and a source-drain bias $V_{\rm SD}$ can be applied across the device to result in sequential single electron tunneling. DQD fabrication and characterization details have been described previously \cite{Liu2014,Liu2015,Liu2017}.

The cavity consists of a half-wavelength ($\lambda/2$) Nb coplanar waveguide resonator with a resonance frequency $f_c$ = 7596 MHz and quality factor $Q_{\rm c} = 4300$ \cite{Wallraff2004, Stehlik2016, Liu2017}. Cavity input and output ports (with coupling rates $\kappa_{\rm in}/2\pi = 0.04$~MHz and $\kappa_{\rm out}/2\pi= 0.8$~MHz) are used to drive the SeSAM with the injection locking tone and to measure the internal field of the maser. The cavity output field is amplified and then characterized using either a spectrum analyzer (R\&S FSV) or heterodyne detection. With heterodyne detection, the output field is demodulated by a local reference tone of frequency $f_{\rm lo}$ to yield the in-phase ($I$) and quadrature-phase ($Q$) components \cite{LiuPRA2015, Stehlik2016}. When the cavity is driven by an injection locking tone, the local reference is always set to the injection locking tone frequency $f_{\rm lo} = f_{\rm in}$ in order to measure the phase $\phi$ of the maser output field relative to the injection locking input tone. 

With $V_{\rm SD}=2$~mV applied, single electron tunneling is allowed when $\epsilon>0$. In this configuration a single electron tunnels down in energy through the device [see Fig.\ 1(a)] and the source-drain bias repumps the DQD to generate the population inversion necessary for photon gain in the cavity \cite{Liu2015, Liu2017}. A trapped charge in the DQD forms an electric dipole moment that interacts with the cavity field with a rate $g_c/2\pi \approx70$~MHz \cite{Delbecq2011, Frey2012, Petersson2012, Delbecq2013, Toida2013, Stehlik2016}. Inelastic interdot tunneling results in a combination of  phonon and photon emission \cite{Wiel2002, Fujisawa1998, Liu2015}. The gain mechanism of the SeSAM is similar to the single emitter limit of a quantum cascade laser, where a macroscopic number of electrons flow through quantum well layers and lead to cascaded photon emission \cite{Faist1994}.

The maser is first characterized in free-running mode with $P_{\rm in} = 0$ (no injection locking tone applied). Figure~\ref{Fig: setup}(b) plots the power spectral density of the output radiation, $S(f)$. The emission peak is nicely fit by a Gaussian with a FWHM $\Gamma = 5.6$~kHz that is 300 times narrower than the cavity linewidth $\kappa_{\rm tot}/2\pi = f_c/Q_c = 1.8$~MHz. The emission signal, and its narrow linewidth, are suggestive of an above-threshold maser state. Maser action is confirmed by measuring the statistics of the output field \cite{Liu2015, Liu2017}. Figure~\ref{Fig: setup}(c) shows the two-dimensional histogram resulting from $1.7\times10^7$ individual $(I,Q)$ measurements that were sampled at a rate of 12.3 MHz. Here $f_{\rm lo} = f_{\rm e} =7595.8$ MHz, where $f_{\rm e}$ is the emission frequency. The $IQ$ histogram has donut shape that is consistent with an above-threshold maser. However, the histogram clearly shows that the phase of the maser output samples all angles in the $(I,Q)$ plane, which indicates there are large phase fluctuations in free running mode. The randomization of phase is attributed to charge noise, which leads to random fluctuations in $\epsilon$ \cite{LiuPRA2015}. In this paper we use injection locking to further improve the output characteristics of the SeSAM.

\section{Injection Locking Results}

We now investigate the degree to which the output characteristics of the SeSAM can be improved using injection locking. In Section III.A we present results showing that the maser emission can be phase locked by driving the input port of the cavity with an injection locking tone. In the injection locked state, the maser output field has a phase noise $\mathcal{L}$ = -99~dBc/Hz at $f_{\rm e}  = 7595.8$ MHz (1.3 MHz offset). In Section III.B, we measure the phase of the maser output field relative to the injection locking input tone as a function of input frequency, and show that it is in good agreement with Adler's prediction. We then measure the injection locking range as a function of injection locking input tone power in Section III.C. The phase and frequency locking range measurements are consistent with each other, giving further evidence that the frequency locking observed in previous work is due to phase stabilization via the injection locking effect \cite{LiuPRA2015}.

\subsection{Phase Locking the SeSAM}

\begin{figure*}[t]
	\begin{center}
		\includegraphics[width=2\columnwidth]{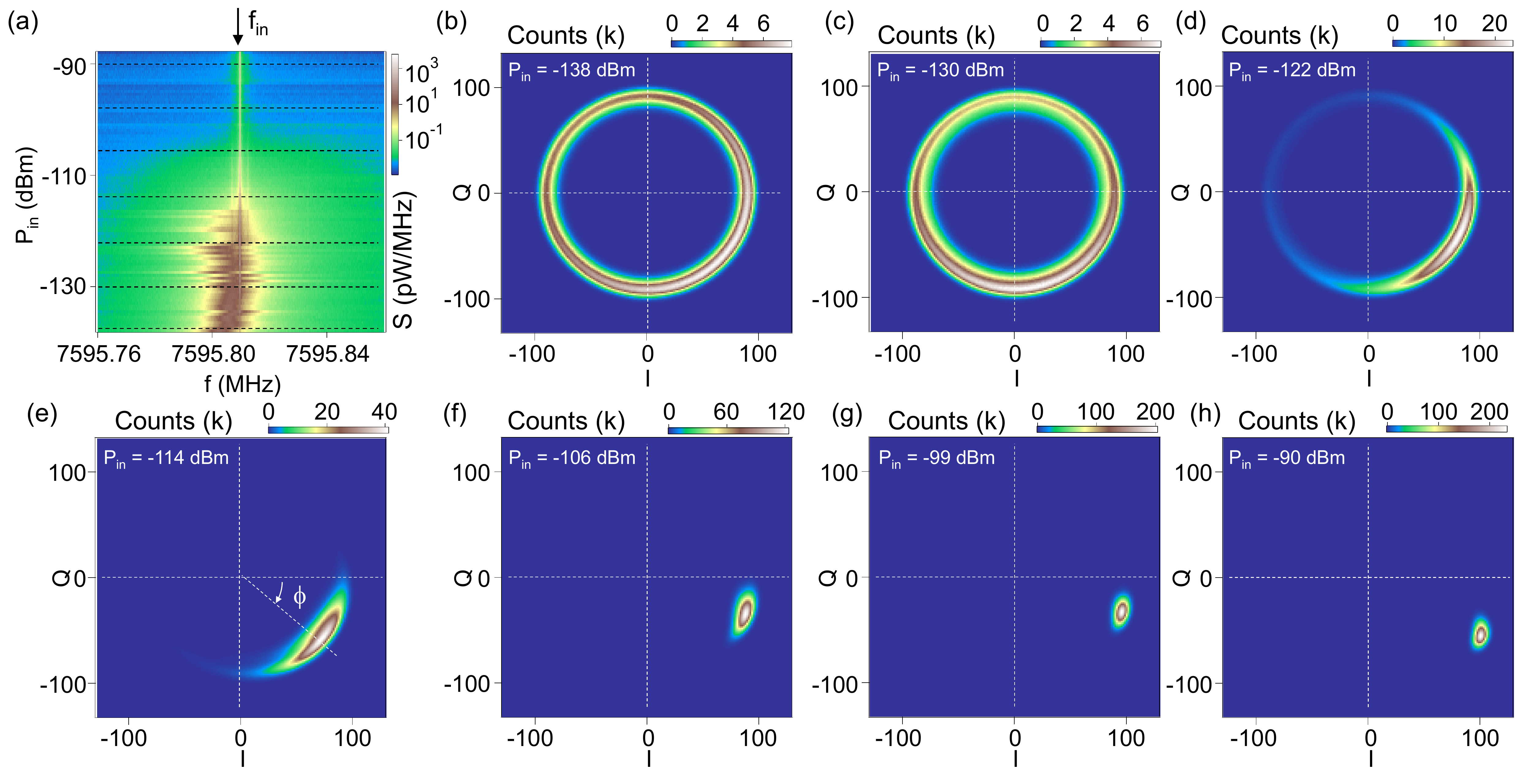}
		\caption{\label{Fig: phase locking} (a) Emission power spectral density $S(f)$ as a function of $P_{\rm in}$. The injection locking tone input frequency $f_{\rm in}$ = 7595.805 MHz is set to be close to the free running maser emission frequency $f_{\rm e} = 7595.805 \pm 0.005$ MHz. Note the significant fluctuations in $f_{\rm e}$ for $P_{\rm in} < $ -125 dBm. The maser linewidth narrows with increasing $P_{\rm in}$ due to injection locking. (b--h) Evolution of the emission statistics in phasor space for $P_{\rm in}$ corresponding to the dashed lines in (a). (b--d) For $P_{\rm in} < -115$~dBm, the cavity field is a combination of the free running maser emission and the cavity input tone at $f_{\rm in}$. In this configuration the maser is unlocked and its phase fluctuates relative to the injection locking input tone. Here the $IQ$ histograms have population in all $360^\circ$ degrees of the $IQ$-plane. (e--h) For larger $P_{\rm in} > -115$~dBm, the maser is phase-locked to the input tone, and the phase distribution is further narrowed with increasing $P_{\rm in}$.}
	\end{center}
\end{figure*}

We first demonstrate frequency narrowing of the maser emission relative to the free-running state using injection locking \cite{LiuPRA2015}. Figure~\ref{Fig: phase locking}(a) shows $S(f)$ as a function of the injection locking input tone power $P_{\rm in}$ with $f_{\rm in}$ = 7595.805~MHz set near the free running emission frequency $f_{\rm e}$ for this device tuning configuration. For negligible input powers ($P_{\rm in}<-125$ dBm) the emission spectrum exhibits a broad peak near $7595.805$ MHz with a typical FWHM $\Gamma \approx 6$~kHz. Due to low frequency charge noise, the center frequency of the free-running emission peak fluctuates within the range $f_{\rm e} = 7595.805 \pm 0.005$ MHz. With $P_{\rm in}$ $>$ -125 dBm, the broad tails of the emission peak are suppressed and the spectrum begins to narrow. The SeSAM eventually locks to the injection locking input tone around $P_{\rm in}=$ -115 dBm. In the injection locked state, the large fluctuations in $f_{\rm e}$ are suppressed and the measured linewidth is $\Gamma$ $\approx$ 100 Hz, more than a factor of 50 narrower than the free-running case \cite{RBW}.

The $IQ$ histograms in Figs.~\ref{Fig: phase locking}(b--h) demonstrate the evolution of the maser output phase relative to the injection locking input tone as $P_{\rm in}$ is increased (for these data sets $f_{\rm lo} = f_{\rm in}$). A movie showing the evolution with $P_{\rm in}$ is included in the supplemental material \cite{SOM}. For small $P_{\rm in}<-120$ dBm, the histograms shown in Fig.~\ref{Fig: phase locking}(b-d) have a ring shape. In contrast to the free-running histogram shown in Fig.\ 1(c), these histograms have an unequal weighting in the $IQ$ plane. For example, the Fig.\ 2(d) histogram has a higher count density for phase angles around $\phi$ = -30$^\circ$. The ring shape indicates that the relative phase of the injection locking input tone and the maser emission are unlocked, while the increased number of counts near a specific phase angle $\phi$ is due to stimulated emission at $f_{\rm in}$. The radius of the rings in the $IQ$-plane doesn't significantly change as $P_{\rm in}$ is increased, which indicates that the total output power of the SeSAM is nearly constant and limited by the DQD photon emission rate. As $P_{\rm in}$ is further increased, the phase distribution continues to narrow, consistent with the narrowing of the emission peak shown in Fig.~\ref{Fig: phase locking}(a) \cite{LiuPRA2015}.

Around $P_{\rm in}= -115$~dBm the ring shaped $IQ$ histogram evolves into a distribution that is localized within a relative phase $\phi \pm \Delta\phi = \phi \pm 3\sigma_{\phi, h} = -40 \pm 60^\circ $, as demonstrated in Fig.~\ref{Fig: phase locking}(e). Here $\phi =\arctan(\bar{I},\bar{Q})$ is the maximumly populated angle and $\sigma_{\phi, h}$ is the measured standard deviation. In this configuration the phase of the maser output is locked to the injection locking input tone. The distribution in phase space is further narrowed with increasing $P_{\rm in}$ as demonstrated by Figs.~\ref{Fig: phase locking}(f--h), where the relative phase is $\phi = -20 \pm 12^\circ$ for $P_{\rm in} > -100$~dBm. The $P_{\rm in}$ value at which phase stabilization occurs is in good agreement with the value of $P_{\rm in}$ where frequency locking occurs, as demonstrated in Fig.~\ref{Fig: phase locking}(a). 

The detected phase fluctuations in the histograms have a standard deviation $\sigma_{\phi,h} = 4^\circ$ for $P_{\rm in} > -100$~dBm. These fluctuations have a contribution from the intrinsic maser output fluctuations with a standard deviation $\sigma_{\phi,0}$ and a contribution from amplifier background noise  $h_{\rm amp}$ [see Fig.\ 1(a)], which has $\langle h_{\rm amp}^\dagger h_{\rm amp} \rangle = 42$  \cite{EichlerPRA2012, Liu2015, Stehlik2016}. The detected field $\alpha = I + i Q$ consists of $\alpha = \alpha_0 + h_{\rm amp}$, where $\alpha_0 = I_0 + i Q_0$ is the cavity output. Given $\alpha_0$ is independent of $h_{\rm amp}$ and $\langle h_{\rm amp} \rangle = 0$, the distribution in the detected phase $\phi_h = \arg(\alpha) =  \arctan({I},{Q})$ (in units of rad) has a standard deviation \[\sigma^2_{\phi,h} = \sigma^2_{\phi,0} + \langle h_{\rm amp}^\dagger h_{\rm amp} \rangle/\left\langle I^2 + Q^2 \right\rangle.\] After subtracting $h_{\rm amp}$, the maser output phase fluctuations have a standard deviation $\sigma_{\phi,0} = 1.5^\circ$ within our detection resolution bandwidth $RBW = 2.6$ MHz. The average phase noise of the locked maser output near $f_{\rm e}$ is then estimated to be $\mathcal{L} = \sigma^2_{\phi,0}/2RBW = 1.3\times10^{-10}$~$\rm rad^2/Hz$ or, equivalently, $\mathcal{L}$  = -99~dBc/Hz at a frequency offset of 1.3 MHz, when $P_{\rm in}>-100$~dBm. For comparison, the phase noise is 40--50 dBc/Hz larger than a typical precision microwave source such as the Keysight E8267D. 

\subsection{Phase Evolution Across the Injection Locking Range}

\begin{figure}[t]
	\begin{center}
		\includegraphics[width=\columnwidth]{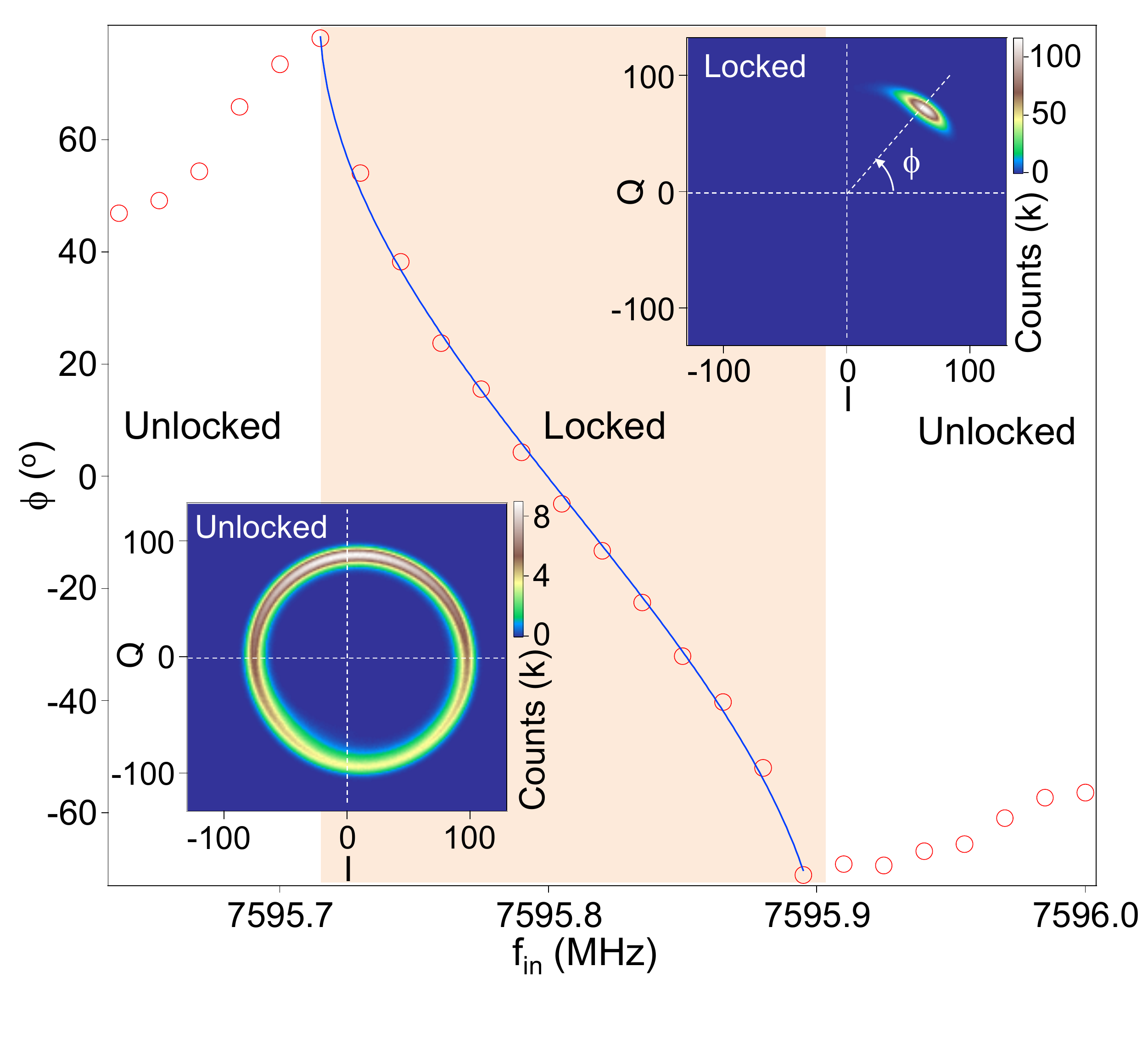}
		\caption{\label{Fig: phase locking range} The phase $\phi$ of the maser emission relative to the injection locking input tone as a function of $f_{\rm in}$ at $P_{\rm in} =-98$~dBm. The blue curve is the prediction from Adler's theory. Left inset: $IQ$ histogram of the output field with $f_{\rm in}=7595.64$ MHz, where the emission phase is unlocked. Right inset: $IQ$ histogram of the maser output field at $f_{\rm in} = 7595.73$ MHz, where the phase of the maser is locked to the injection locking input tone with $\phi=48^\circ$.}
	\end{center}	
\end{figure}

We now investigate the relative phase $\phi$ between the maser output and the injection locking input tone across the full injection locking range. The insets of Fig.~\ref{Fig: phase locking range} show $IQ$ histograms acquired with $P_{\rm in}=-98$~dBm at $f_{\rm in}=7595.64$ MHz (left inset) and $f_{\rm in}=7595.73$ MHz (right inset). With $f_{\rm in}=7595.64$ MHz, which is detuned by 0.17 MHz from the free running maser frequency $f_{\rm e} = 7595.81$, the $IQ$ distribution has a ring-like shape and thus the phase is unlocked. Note that in this regime the output is essentially the sum of two different tones, and this results in a noticeable offset in the ring. When $f_{\rm in}$ approaches $f_{\rm e}$, the phase will be localized within a small range, as demonstrated in the right inset, which shows a distribution that is limited to $\phi= 48\pm 15^\circ$. Here $f_{\rm in}$ is detuned from $f_{\rm e}$ by only 0.08 MHz. 

The main panel of Fig.~\ref{Fig: phase locking range} shows $\phi$ as a function of $f_{\rm in}$ with $P_{\rm in} =-98$~dBm. Within the indicated frequency range of $\Delta f_{\rm in} = 0.19$~MHz, the histograms are similar to the right inset and show output phases in the range $\phi \in\left(-90^\circ, 90^\circ \right)$. The maser output is thus ``phase locked" to the input tone when $|f_{\rm in}-f_{\rm e}|$ is small. 

The measured phase can be compared with predictions from Adler's theory, which analyzes the maser dynamics when the injection locking tone input power is small compared to the free running emission power \cite{Adler1946}.  We express the cavity output field in the lab frame as
\begin{equation}
\alpha(t) = I(t) + iQ(t) = \sqrt{\frac{P_{\rm out}}{RBW\; hf_e}} e^{2\pi if_{\rm in}t + i\phi(t)}
\end{equation}
where $P_{\rm out}$ is the output power. The relative phase follows the Adler equation:
\begin{equation}
\frac{d\phi}{dt}+2\pi(f_{\rm in}-f_{\rm e}) =-2\pi\frac{\Delta f_{\rm in}}{2}\sin(\phi).
\label{Eq: Adler equation}
\end{equation}
In the injection locking range $|f_{\rm in}-f_{\rm e}|<\Delta f_{\rm in}/2$, Eq.\ (\ref{Eq: Adler equation}) has a static solution 
\begin{equation}
	\phi=\arcsin \left[2(f_{\rm e}-f_{\rm in})/\Delta f_{\rm in}\right].
	\label{Eq: Adler phase}
\end{equation} 
Fluctuations in $\phi$ can be introduced by fluctuations in $f_{\rm e}$ and the intrinsic standard deviation $\sigma_{\phi,0}$ diverges near the boundaries of the injection locking range. Outside of this range $\phi$ is unlocked. The phase dependence predicted by the Adler equation is plotted as the blue curve in Fig.~\ref{Fig: phase locking range} and is in good agreement with our data.

\subsection{Injection Locking Range}

\begin{figure}[t]
	\begin{center}
		\includegraphics[width=\columnwidth]{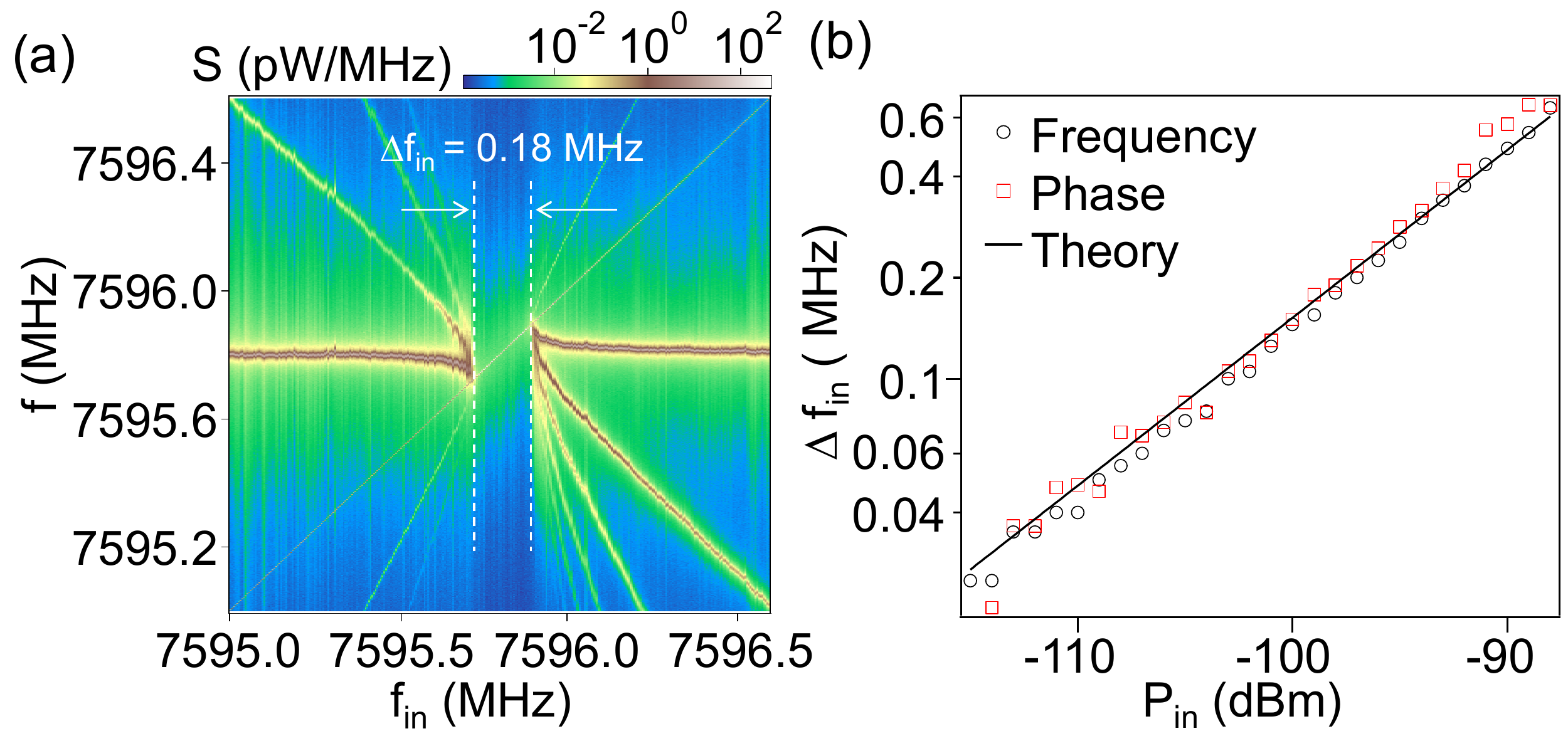}
		\caption{\label{Fig: frequency locking range} (a) $S(f)$ measured as a function of $f_{\rm in}$ with $P_{\rm in} = -98$ dBm. The white dashed lines indicate the injection locking range $\Delta f_{\rm in}$. (b) $\Delta f_{\rm in}$ as a function of  $P_{\rm in}$. The locking range is extracted from measurements of $S(f)$ (circles) and from the phase locking range in the $IQ$ histograms (squares). The black line is a fit to the power law $\Delta f_{\rm in} \propto \sqrt{P_{\rm in}}$ prediction of the Adler equation.}
	\end{center}
	\vspace{-0.4cm}
\end{figure}

We next determine the frequency locking range from measurements of $S(f)$ and compare these data with the phase locking measurements presented in the previous section. Figure~\ref{Fig: frequency locking range}(a) shows a color-scale plot of $S(f)$ as a function of $f_{\rm in}$ measured with $P_{\rm in}$ = -98 dBm. Similar to our previous work \cite{LiuPRA2015}, the input tone has little effect on the maser emission when $f_{\rm in}$ is far-detuned from $f_{\rm e}$. As $f_{\rm in}$ approaches $f_{\rm e}$, frequency pulling is visible and emission sidebands appear as a mixing between the injection locking input tone and the free running maser emission \cite{Siegman1986, Armand1969, LiuPRA2015}. The maser then abruptly locks to $f_{\rm in}$, and remains locked to $f_{\rm in}$ over a frequency range $\Delta f_{\rm in}$ = 0.18 MHz. The frequency locking range is consistent with the phase locking data shown in Fig.~\ref{Fig: phase locking range}, which is measured at the same $P_{\rm in}$. 

By repeating these measurements at different $P_{\rm in}$, we obtain the data shown in Fig.~\ref{Fig: frequency locking range}(b), where $\Delta f_{\rm in}$ measured by the two methods is plotted as a function of $P_{\rm in}$. The measurements are in good agreement, verifying that the frequency locking we observe in measurements of $S(f)$ is due to the injection locking effect \cite{LiuPRA2015}. 
The black line in Fig.~\ref{Fig: frequency locking range}(b) is a fit to the power law relation $\Delta f_{\rm in}$ = $A_{\rm M}$ $\sqrt{P_{\rm in}}$, with the measured prefactor $A_{\rm M}$ = $ (0.48 \pm 0.16) \times 10 ^6\; {\rm MHz/\sqrt{W}}$, where the error bar is due to the uncertainty in the input transmission line losses. From theory, we find: \begin{equation*}
	A_{\rm T} = \frac{C_\kappa}{\sqrt{P_{\rm out}}} \frac{\kappa_{\rm tot}}{2\pi} = \left(0.87 \pm 0.29 \right) \times 10 ^6\; {\rm MHz/\sqrt{W}},
\end{equation*} where the cavity prefactor $C_\kappa =  2\sqrt{\kappa_{\rm in}\kappa_{\rm out}}/\kappa_{\rm tot}$ accounts for internal cavity losses and is obtained using cavity input-output theory \cite{Siegman1986, LiuPRA2015}. The error bar is due to the uncertainty in $\kappa_{\rm in/out}$ and the calibration of $P_{\rm out}$. We therefore find reasonable agreement between the data and the predictions from Adler's theory, considering the uncertainties in the transmission line losses.

\section{Microwave Frequency Comb}

\begin{figure}
	\begin{center}
		\includegraphics[width=\columnwidth]{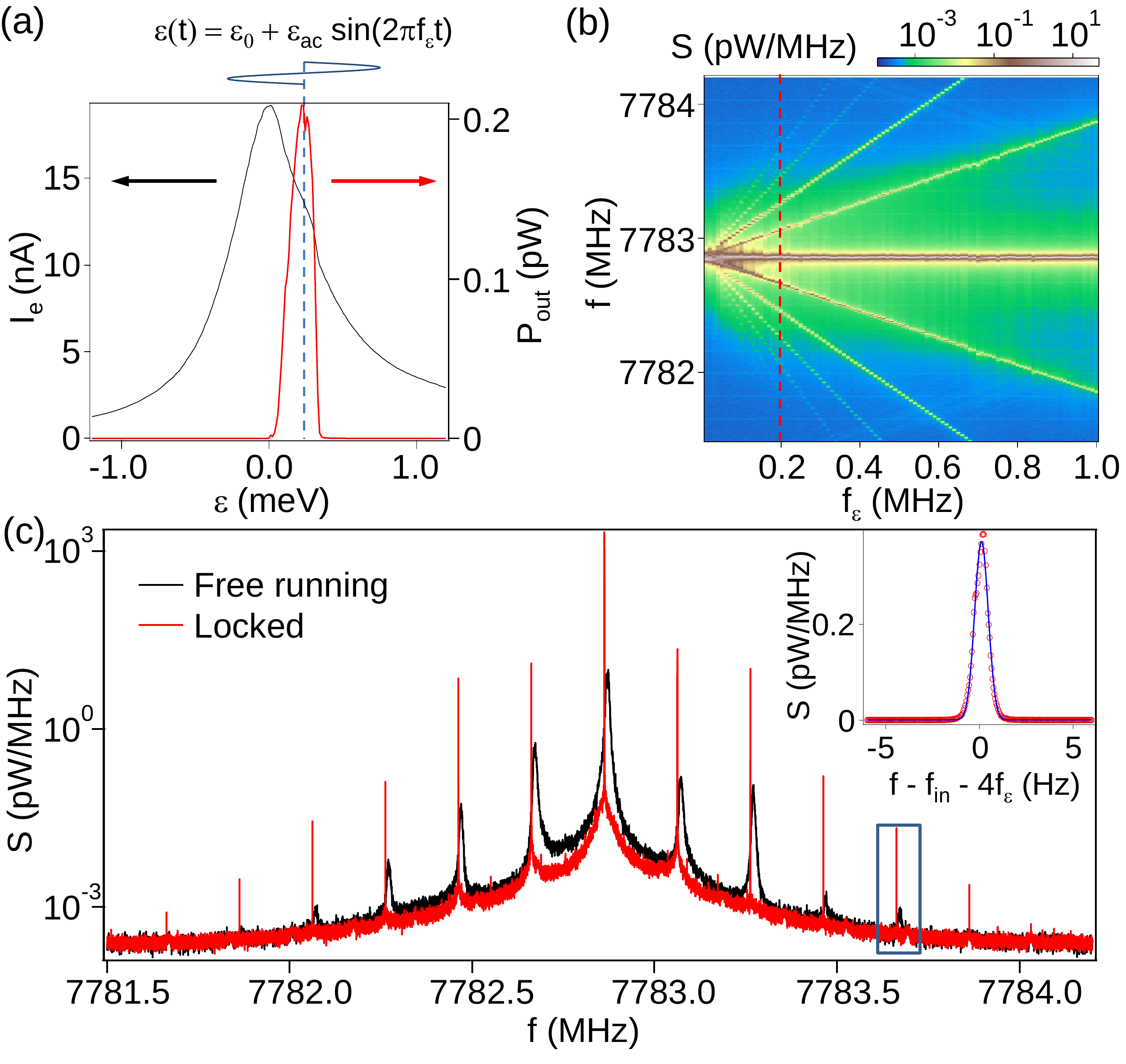}
		\caption{\label{Fig: frequency comb} (a) $I_e$ (black) and $P_{\rm out}$ (red) plotted as a function of $\epsilon$. A sinusoidal drive is applied to the detuning parameter to modulate the gain of the DQD maser. (b) $S(f)$ as a function of $f_\epsilon$ with $\epsilon_{\rm ac}=0.2$ meV and $P_{\rm in}=0$. A frequency comb is observed with the spacing set by $f_\epsilon$. (c) $S(f)$ with $f_\epsilon=0.2$~MHz and $\epsilon_{\rm ac}=0.2$~meV. The black curve is the frequency comb measured in free running mode with $P_{\rm in} = 0$. The red curve is the frequency comb acquired under injection locking conditions with $f_{\rm in} = 7782.9$ MHz and $P_{\rm in} = -108$ dBm. Inset: A zoom-in of the injection locked comb at the fourth sideband.}
	\end{center}
	\vspace{-0.4cm}
\end{figure}

We have so far examined the output characteristics of the SeSAM in free-running mode and under the influence of an injection locking tone. In this section, we investigate the output characteristics of the SeSAM while a periodic modulation is applied to the DQD energy levels, which modulates the gain medium. With the periodic modulation applied, we observe a comb-like emission spectrum, where the spacing between the emission peaks is set by the modulation frequency. The SeSAM frequency comb can also be operated under injection locking conditions, which leads to a dramatic narrowing of the emission peaks. The data presented in this section were acquired on a different device that has an emission frequency $f_{\rm e} = 7782.86$ MHz and linewidth $\Gamma = 3$~kHz.

The modulation method is described in Fig.\ \ref{Fig: frequency comb}(a), which plots the electron current $I_e$ and $P_{\rm out}$ as a function of $\epsilon$. In free-running mode, the maximum output power $P_{\rm out} = 0.2$ pW is obtained at an offset detuning $\epsilon_0 = 0.2$ meV due to a strong phonon sideband \cite{Gullans2015}. We next modulate the gain medium by applying a sine wave to the DQD gates, such that $\epsilon = \epsilon_0 + \epsilon_{\rm ac}\sin(2\pi f_\epsilon t)$. Here $\epsilon_{\rm ac}$ and $f_\epsilon$ are the amplitude and frequency of the detuning modulation. As shown in Fig.~\ref{Fig: frequency comb}(a), the SeSAM emission power is strongly detuning dependent. Therefore the effective gain rate will be modulated by the sinusoidal gate drive \cite{Liu2015}.

Figure~\ref{Fig: frequency comb}(b) plots $S(f)$ as a function of $f_\epsilon$ with $\epsilon_{\rm ac}=0.2$~meV. We observe a central emission peak around $f$ = 7782.86 MHz that is independent of $f_\epsilon$. In addition to the central emission peak we observe a series of narrow emission peaks that shift away from the central emission peak as $f_\epsilon$ is increased. Up to 4 emission sidebands are clearly observed on both the low and high frequency sides of the central emission peak. With such a large modulation amplitude applied, photoemission from the DQD will turn on and off at a beat frequency $f_\epsilon$. The beating in the time domain results in sidebands in the frequency domain, with a sideband spacing set by $f_\epsilon$. The black curve in Fig.~\ref{Fig: frequency comb}(c) shows a line cut through the data in Fig.~\ref{Fig: frequency comb}(b) at $f_\epsilon=0.2$~MHz. The sidebands can be fit to a Lorentzian with a linewidth of 8 kHz, similar to the free-running maser linewidth $\Gamma$ = 3 kHz. 

The linewidth of the emission peaks in the frequency comb can be significantly improved using the injection locking effect \cite{Diddams2000}. For example, the red curve in Fig.~\ref{Fig: frequency comb}(c) shows $S(f)$ when the frequency comb is injection locked to an input tone at $f_{\rm in} = 7782.86$ MHz and $P_{\rm in} = -108$ dBm. Compared to the free-running frequency comb data, the peak height and linewidth of the injection locked frequency comb have been dramatically improved. In addition, we observe 2 additional sidebands on both the low and high frequency sides of the central emission peak. The inset of Fig.~\ref{Fig: frequency comb}(c) shows $S(f)$ measured near the fourth sideband on the high frequency side of the central emission peak [near $f$ = 7783.65  MHz, see rectangle in main panel of Fig.\ 5(c)]. The sideband  is best fit to a Gaussian of width 0.9 Hz, which is most likely limited by the 1 Hz resolution bandwidth of the microwave frequency spectrum analyzer \cite{RBW}.

\section{Conclusion and Outlook}

We have presented experimental evidence of phase locking of a semiconductor DQD single atom maser (SeSAM). The statistics of the maser emission in the complex plane demonstrates that the SeSAM can be phase locked to an injection locking input tone resulting in a emission signal with a phase noise $\mathcal{L}$ = -99~dBc/Hz at a frequency offset of 1.3 MHz. Both phase and frequency locking data are shown to be in good agreement with Adler's prediction. In addition, we utilize the electrical tunability of the DQD energy level structure to modulate the DQD gain medium. The resulting emission spectrum is a frequency comb, where individual emission peaks in the comb have a linewidth of around 8~kHz. By injection locking the SeSAM, we reach linewidths $<$ 1 Hz, an 8000-fold improvement. The SeSAM allows for studies of fundamental light-matter interactions in condensed matter systems. These measurements demonstrate that a single DQD may serve as a compact low temperature microwave source that is suitable for use in quantum computing experiments.

\begin{acknowledgments}
We thank M. J. Gullans and J. M. Taylor for helpful discussions and acknowledge support from the Packard Foundation, the National Science Foundation Grant No.\ DMR-1409556, and the Gordon and Betty Moore Foundation’s EPiQS Initiative through Grant GBMF4535. Devices were fabricated in the Princeton University Quantum Device Nanofabrication Laboratory.
\end{acknowledgments}

\bibliographystyle{apsrev_lyy2017}
\bibliography{SAM_locking_v7}

\begin{thebibliography}{41}
\bibitem[{RBW()}]{RBW}
\bibinfo{note}{The frequency resolution is limited by the resolution bandwidth
  of the R\&S spectrum analyzer FSV. For Fig.~2 the $RBW$ is set to 0.1 kHz, for
  Fig.~4 the $RBW$ is set to 1 kHz, for Fig.~5(b) and Fig.~5(c) main panel the
  $RBW$ is set to 0.1 kHz, and for Fig.~5(c) inset and Fig.~5(d) the $RBW$ is set
  to 1 Hz.}
\end{thebibliography}

\end{document}